\documentclass[9pt,twocolumn,twoside]{osajnl}

\journal{optica} 

\setboolean{shortarticle}{false}
\usepackage{arydshln,mathtools}
\usepackage{epsfig,amsmath,amsfonts,stackrel,bm,mathrsfs}

\newcommand{\expb}[1]{\exp\left\{ #1 \right\}}

\newcommand{\sst}[2]{{#1}_{\text{#2}}}

\newcommand\balphlist[1]{\newcounter{#1}
\begin{list}
  {{\bf \alph{#1})}}   
  {\usecounter{#1}\setlength{\rightmargin}{0cm}
   \itemsep 1ex\parsep 1ex}}
\newcommand\alphlist[1]{\newcounter{#1}
\begin{list}
  {{\alph{#1})}}   
  {\usecounter{#1}\setlength{\rightmargin}{0cm}
   \itemsep 1ex\parsep 1ex}}
\newcommand\bnumlist[1]{\newcounter{#1}
\begin{list}
  {{\bf \arabic{#1})}}   
  {\usecounter{#1}\setlength{\rightmargin}{0cm}
   \itemsep 1ex\parsep 1ex}}
\newcommand\numlist[1]{\newcounter{#1}
\begin{list}
  {{\arabic{#1})}}   
  {\usecounter{#1}\setlength{\rightmargin}{0cm}
   \itemsep 1ex\parsep 1ex}}
\newcommand\romlist[1]{\newcounter{#1}
\begin{list}
  {{(\roman{#1})}}   
  {\usecounter{#1}\setlength{\rightmargin}{0cm}
   \itemsep 1ex\parsep 1ex}}
\newcommand\bromlist[1]{\newcounter{#1}
\begin{list}
  {{\bf \roman{#1})}}   
  {\usecounter{#1}\setlength{\rightmargin}{0cm}
   \itemsep 1ex\parsep 1ex}}
\newcommand\problist{\newcounter{probc}
\begin{list}
  {{\bf \arabic{probc}.}}   
  {\usecounter{probc}\setlength{\rightmargin}{0cm}
   \itemsep 1ex\parsep 1ex}}
\newcommand\problistn[1]{\newcounter{#1}
\begin{list}
  {{\bf \arabic{#1}.}}   
  {\usecounter{#1}\setlength{\rightmargin}{0cm}
   \itemsep 1ex\parsep 1ex}}
\newcommand\subproblist[1]{\newcounter{#1}
\begin{list}
  {{\bf \arabic{probc}.\alph{#1})}}   
  {\usecounter{#1}\setlength{\rightmargin}{0cm}
   \itemsep 1ex\parsep 1ex}}

\title{Limited-angle tomographic reconstruction of dense layered objects by dynamical machine learning}

\author[1,*]{Iksung Kang}
\author[2,$\dagger$]{Alexandre Goy}
\author[2,3]{George Barbastathis}

\affil[1]{Department of Electrical Engineering and Computer Science, Massachusetts Institute of Technology, 77 Massachusetts Ave, Cambridge, MA 02139, USA}
\affil[2]{Department of Mechanical Engineering, Massachusetts Institute of Technology, 77 Massachusetts Ave, Cambridge, MA 02139, USA}
\affil[3]{Singapore-MIT Alliance for Research and Technology (SMART) Centre, 1 Create Way, Singapore 117543, Singapore}
\affil[*]{Corresponding author: iskang@mit.edu}
\affil[$\dagger$]{Present address: Omnisens SA, 1110 Morges, Switzerland}

\begin{abstract}
Limited-angle tomography of strongly scattering quasi-transparent objects is a challenging, highly ill-posed problem with practical implications in medical and biological imaging, manufacturing, automation, and environmental and food security. Regularizing priors are necessary to reduce artifacts by improving the condition of such problems. Recently, it was shown that one effective way to learn the priors for strongly scattering yet highly structured 3D objects, e.g. layered and Manhattan, is by a static neural network\ [Goy \textit{et al}, \textit{Proc. Natl. Acad. Sci.} 116, 19848-19856 (2019)]. Here, we present a radically different approach where the collection of raw images from multiple angles is viewed analogously to a dynamical system driven by the object-dependent forward scattering operator. The sequence index in angle of illumination plays the role of discrete time in the dynamical system analogy. Thus, the imaging problem turns into a problem of nonlinear system identification, which also suggests dynamical learning as better fit to regularize the reconstructions. We devised a recurrent neural network (RNN) architecture with a novel split-convolutional gated recurrent unit (SC-GRU) as the fundamental building block. Through comprehensive comparison of several quantitative metrics, we show that the dynamic method improves upon previous static approaches with fewer artifacts and better overall reconstruction fidelity.
\end{abstract}

\setboolean{displaycopyright}{true}

\begin{document}

\maketitle

\section{Introduction}
\begin{figure*}[t!]
    \centering
    \includegraphics[width=0.75\textwidth]{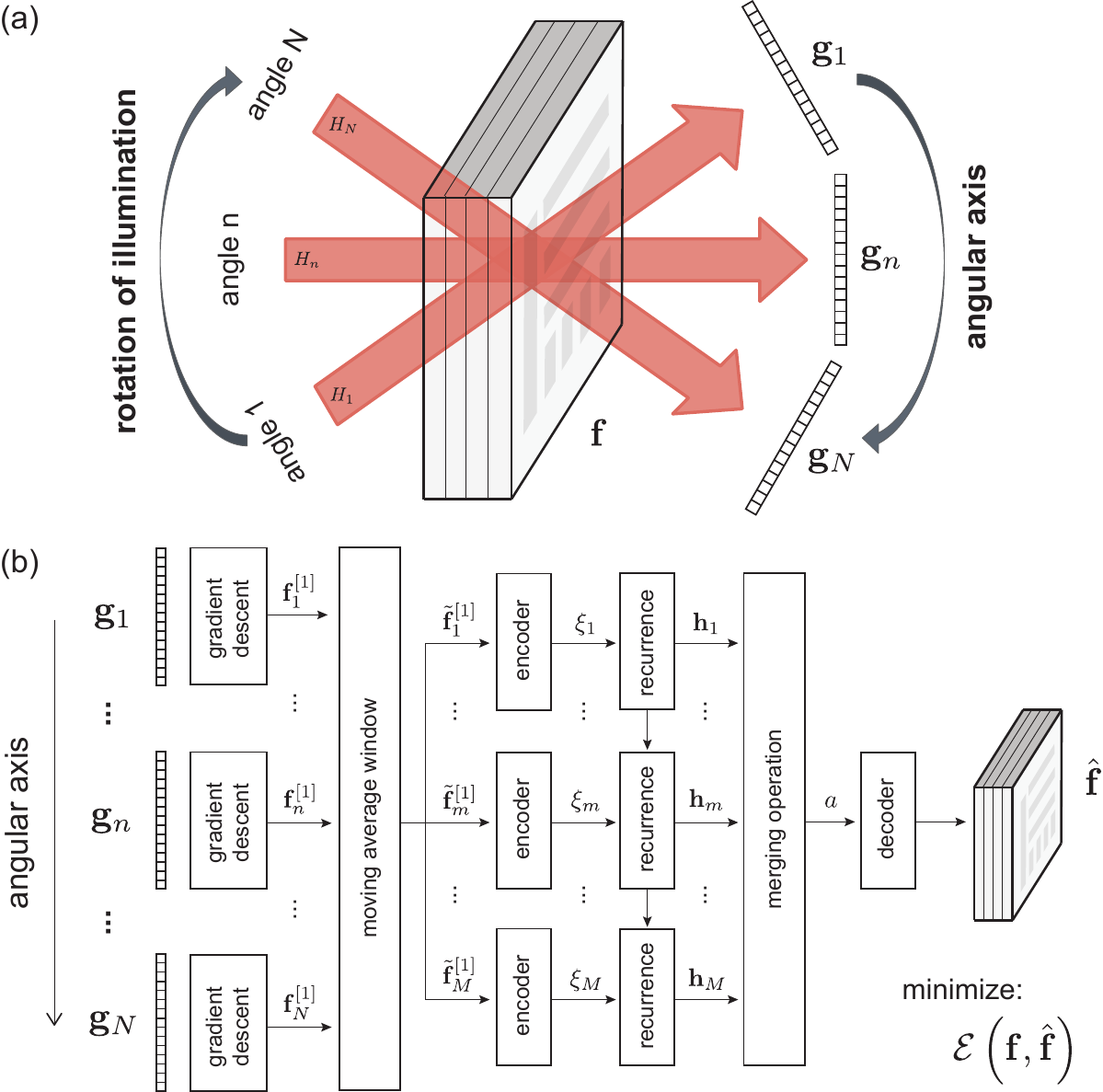}
    \caption{(a) Each angle of illumination, here labelled as angular axis, corresponds to a time step in an analogous temporal axis. (b) The raw intensity diffraction pattern $\mathbf{g}_n,\: n\!=\!1,\ldots,N\!\!=\!\!42$\ \ of the at $n$-th angular sequence step is followed by gradient descent and moving average operations to construct a shorter Approximant sequence $\mathbf{\tilde{f}}_m{}^{[1]},\: m\!=\!1,\ldots,M\!\!=\!\!12$. The Approximants $\mathbf{\tilde{f}}_m{}^{[1]}$ are encoded to $\xi_m$ and fed to the recurrent dynamical operation whose output sequence $\mathbf{h}_m, m\!=\!1,\ldots,12$\ \ the angular attention scheme merges into a single representation $a$, and that is finally decoded to produce the 3D reconstruction $\mathbf{\hat{f}}$. Training adapts the weights of the learned operators in this architecture to minimize the training loss function $\mathcal{E}(\mathbf{f},\hat{\mathbf{f}})$ between $\mathbf{\hat{f}}$ and the ground truth object $\mathbf{f}$.}
    \label{fig:introduction}
\end{figure*}

Optical tomography reconstructs the three-dimensional (3D) internal refractive index profile by illuminating the sample at several angles and processing the respective raw intensity images. The reconstruction scheme depends on the scattering model that is appropriate for a given situation. If the rays through the sample can be well approximated as straight lines, then accumulation of absorption and phase delay along the rays is an adequate forward model, {\it i.e.} the projection or Radon transform approximation applies. This is often the case with hard x-rays through most materials including biological tissue; for that reason, Radon transform inversion has been widely studied\ \cite{radon1986determination,radon1917determination,bracewell1967inversion,feldkamp1984practical,dreike1976convolution,wang1993general,kudo1991helical,grangeat1991mathematical,katsevich2002analysis,choi2007tomographic}.

The next level of complexity arises when diffraction and multiple scattering must be taken into account in the forward model; then, the Born or Rytov expansions and the Lippmann-Schwinger integral equation \cite{ishimaru2017electromagnetic,tatarski2016wave,wolf1969three,devaney1981inverse,pham2020three} are more appropriate. These follow from the scalar Helmholtz equation using different forms of expansion for the scattered field \cite{marks2006family}. In all these approaches, weak scattering is obtained from the first order in the series expansion. Holographic approaches to volumetric reconstruction generally rely on this first expansion term\ \cite{milgram2002computational,tian2010quantitative,hahn2008wide,park2009recent,nehmetallah2012applications,williams2013digital,brady2009compressive,choi2010compressive,rivenson2018phase,wu2019bright,rivenson2019deep,zhang2018twin}. Often, solving the Lippmann-Schwinger equation is the most robust approach to account for multiple scattering, but even then the solution is iterative and requires excessive amount of computation especially for complex 3D geometries. The inversion of these forward models to obtain the refractive index in 3D is referred to as inverse scattering, also a well studied topic \cite{kamilov2016recursive,kamilov2016optical,giorgi2013application,chew1990reconstruction,sun2018efficient,lu1985multidimensional,lu1986jkm,tsihrintzis2000higher}.

An alternative to the integral methods is the beam propagation method (BPM), which sections the sample along the propagation distance $z$ into slices, each slice scattering according to the thin transparency model, and propagates the field from one slice to the next through the object\ \cite{feit1980computation}. Despite some compromise in accuracy, BPM offers comparatively light load of computation and has been used as forward model for 3D reconstructions\ \cite{pham2020three}. The analogy of the BPM computational structure with a neural network was exploited, in conjunction with gradient descent optimization, to obtain the 3D refractive index as the ``weights'' of the analogous neural network in the learning tomography approach \cite{kamilov2015learning,shoreh2017optical,lim2018learning}. BPM has also been used with more traditional sparsity-based inverse methods\ \cite{kamilov2016optical,chowdhury2019high}. Later, a machine learning approach with a convolutional neural network (CNN) replacing the iterative gradient descent algorithm exhibited even better robustness to strong scattering for layered objects, which match well with the BPM assumptions \cite{goy2019high}. Despite great progress reported by these prior works, the problem of reconstruction through multiple scattering remains difficult due to the extreme ill-posedness and uncertainty in the forward operator; residual distortion and artifacts are not uncommon in experimental reconstructions. 

Inverse scattering, as inverse problems in general, may be approached in a number of different ways to regularize the ill-posedness and thus provide some immunity to noise \cite{bertero1998introduction,candes2006robust}. Recently, thanks to a ground-breaking observation from 2010 that sparsity can be learnt by a deep neural network \cite{gregor2010learning}, the idea of using machine learning to approximate solutions to inverse problems also caught on \cite{barbastathis2019use}. In the context of tomography, in particular, deep neural networks have been used to invert the Radon transform \cite{jin2017deep} and recursive Born model \cite{kamilov2016recursive}, and were also the basis of some of the papers we cited earlier on holographic 3D reconstruction\ \cite{wu2019bright,rivenson2018phase,rivenson2019deep}, learning tomography\ \cite{kamilov2015learning,shoreh2017optical,lim2018learning}, and multi-layered strongly scattering objects\ \cite{goy2019high}. In prior work on tomography using machine learning, generally, the intensity projections are all fed as inputs to a computational architecture that includes a neural network, and the output is the 3D reconstruction of the refractive index. The role of the neural network is to learn the priors that apply to the particular class of objects being considered and the relationship of these priors to the forward operator (Born, BPM, etc.) so as to produce a reasonable estimate of the inverse.

Here we propose a rather distinct approach to exploit machine learning for 3D refractive index reconstruction under strong scattering conditions. Our motivation is that, as the angle of illumination is changed, the light goes through {\em the same scattering volume,} but the scattering events follow a different sequence. At the same time, the intensity diffraction pattern obtained from a new angle of illumination adds information to the tomographic problem, but that information is constrained by ({\it i.e.}, is not orthogonal to) the previously obtained patterns. We interpret this as similar to a dynamical system, where as time evolves and new inputs arrive, the output is constrained by the history of earlier inputs. (The convolution integral is the simplest and best known expression of this relationship between the output of a system and the history of the system's input.)

The analogy between strong scattering tomography and a dynamical system suggests the recurrent neural network (RNN) architecture as a strong candidate to process intensity diffraction patterns in sequence, as they are obtained one after the other; and process them recurrently so that each intensity diffraction pattern from a new angle improves over the reconstructions obtained from the previous angles. Thus, we treat multiple diffraction patterns under different illumination angles as a temporal sequence, as shown in Figure~\ref{fig:introduction}. The angle index $\theta$ replaces what in a dynamical system would have been the time $t$. This idea is intuitively appealing; it also leads to considerable improvement in the reconstructions, removing certain artifacts that were visible in \cite{goy2019high}, as we will show in section~\ref{sec:results}. 

The way we propose to use RNNs in this problem is quite distinct from the recurrent architecture proposed first in \cite{gregor2010learning} and subsequently implemented, replacing the recurrence by a cascade of distinct neural networks, in \cite{jin2017deep,inv:mardani2017a,inv:mardani2017b}, among others. In these prior works, the input to the recurrence can be thought of as clamped to the raw measurement, as in the proximal gradient \cite{inv:daubechies04} and related methods; whereas, in our case, the input to the recurrence is itself dynamic, with the raw intensity diffraction patterns from different angles forming the input sequence. Moreover, by utilizing a modified gated recurrent unit (more on this below) rather than a standard neural network, we do not need to break the recurrence up into a cascade. 

Typical applications of RNNs \cite{williams1989learning,hochreiter1997long} are in temporal sequence learning and identification. In imaging and computer vision, RNN is applied in 2D and 3D: video frame prediction \cite{xingjian2015convolutional,wang2018eidetic,wang2017predrnn,wang2018predrnn++}, depth map prediction \cite{cs2018depthnet}, shape inpainting \cite{wang2017shape}; and stereo reconstruction \cite{liu2020novel,choy20163d} or segmentation \cite{le2017multi,stollenga2015parallel} from multi-view images, respectively. Stereo, in particular, bears certain similarities to our tomographic problem here, as sequential multiple views can be treated as a temporal sequence. To establish the surface shape, the RNNs in these prior works learn to enforce consistency in the raw 2D images from each view and resolve the redundancy between adjacent views in recursive fashion through the time sequence ({\it i.e.}, the sequence of view angles). Non-RNN learning approaches have also been used in stereo, e.g. Gaussian mixture models\ \cite{hou2019multi}.

In this work, we replaced the standard long-short term memory (LSTM)\ \cite{hochreiter1997long} implementation of RNNs with a modified version of the newer gated recurrent unit (GRU) \cite{cho2014learning}. The GRU has the advantage of fewer parameters but generalizes comparably with the LSTM. Our GRU employs a split convolutional scheme to explicitly account for the asymmetry between the lateral and axial axes of propagation, and an angular attention mechanism that learns how to reward specific angles in proportion to their contribution to reconstruction quality. For isotropic (in the ensemble sense) samples as we consider here, it turns out that the attention mechanism treats all angles equally, yet we found that its presence still improves the quality of the training algorithm. For more general sample classes with spatially anisotropic structure, angular attention may be expected to treat different angles of illumination with more disparity.

Details in experiments are delineated in Section~\ref{sec:experiment}. The computational elements are all described in Section~\ref{sec:comput_arch}, while training and testing procedures are illustrated in Section~\ref{sec:train_and_test}. The results of our experimental study are in Section~\ref{sec:results}, showing significant improvement over static neural network-based reconstructions of the same data both visually and in terms of several quantitative metrics. We also include results from an ablation study that indicates the relative significance of the new components we introduced to the quality of the reconstructions.

\section{Experiment} \label{sec:experiment}

\begin{figure}[t!]
    \centering
    \includegraphics[width=\linewidth]{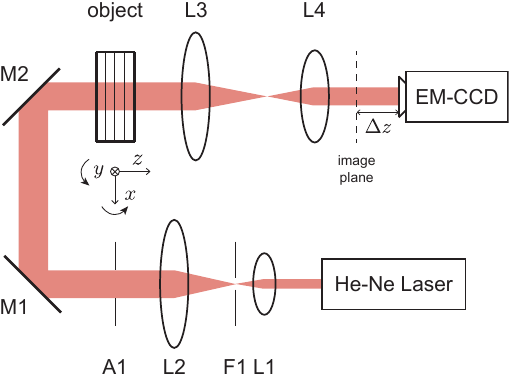}
    \caption{Optical apparatus used for experimental data acquisition\ \protect{\cite{goy2019high}}. L$1-4$: lenses, F$1$: pinhole, A$1$: aperture, EM-CCD: electron-multiplying charge coupled device. $f_{\text{L}_3}:f_{\text{L}_4} = 2:1$. The object is rotated along both $x$ and $y$ axes. The defocus distance between the conjugate plane to the exit object surface and the EM-CCD is $\Delta z = 58.2\:\text{mm}$.}
    \label{fig:optical_apparatus}
\end{figure}

The experimental data are the same as in \cite{goy2019high}, whose experimental apparatus is summarized in Figure~\ref{fig:optical_apparatus}. We repeat the description here for the readers' convenience. The He-Ne laser (Thorlabs HNL210L, power: $20\:\text{mW}$, $\lambda = 632.8\:\text{nm}$) illuminated the sample after spatial filtering and beam expansion. The illumination beam was then de-magnified by the telescope ($f_{\text{L}_3} : f_{\text{L}_4} = 2:1$), and the EM-CCD (Rolera EM-C$2$, pixel pitch: $8\:\mu\text{m}$, acquisition window dimension: $1002\:\times\:1004$) captured the experimental intensity diffraction patterns. The integration time for each frame was $2\:\text{ms}$, and the EM gain was set to $\times 1$. The optical power of the laser was strong enough for the captured intensities to be comfortably outside the shot-noise limited regime.

Each layer of the sample was made of fused silica slabs ($n=1.457$ at $632.8$ nm and at $20\:^\circ$C). Slab thickness was $0.5\text{mm}$, and patterns were carefully etched to the depth of $575\pm 5$ nm on the top surface of each of the four slabs. To reduce the difference between refractive indices, gaps between adjacent layers were filled with oil ($n = 1.4005\pm 0.0002$ at $632.8$ nm and at $20^\circ$C), yielding binary phase depth of $-0.323\pm 0.006\:\text{rad}$. The diffraction patterns used for training were prepared with simulation precisely matched to the apparatus of Figure~\ref{fig:optical_apparatus}. For testing, we used a set of diffraction patterns that was acquired experimentally.

Objects used for both simulation and experiment are dense-layered, transparent, \textit{i.e.} of negligible amplitude modulation, and of binary refractive index. They were drawn from a database of IC layout segments\ \cite{goy2019high}. The feature depth of $575\pm 5\:\text{nm}$ and refractive index contrast $0.0565\pm0.0002$ at $632.8$ nm and at $20\:^\circ$C were such that weak scattering assumptions are invalid and strong scattering has to be necessarily taken into account. The Fresnel number ranged from $0.7$ to $5.5$ for the given defocus amount $\Delta z=58.2\:\text{mm}$ for the range of object feature sizes. 

To implement the raw image acquisition scheme, the sample was rotated from $-10$ degree to $10$ degree with a $1$-degree increment along both the $x$ and $y$ axes, while the illumination beam and detector remained still. This resulted in $N=42$ angles and intensity diffraction patterns in total (see Section~\ref{sec:comput_arch}.\ref{subsec:conv_enc_dec}). Note that \cite{goy2019high} only utilized $22$ patterns out of with a $2$-degree increment along both $x$ and $y$ axes. The comparisons we show later are still fair because we retrained all the algorithms of \cite{goy2019high} for the $42$ angles and $1^\circ$ increment.

\section{Computational architecture}\label{sec:comput_arch}

\begin{figure*}[t!]
    \centering
    \includegraphics[width=\textwidth]{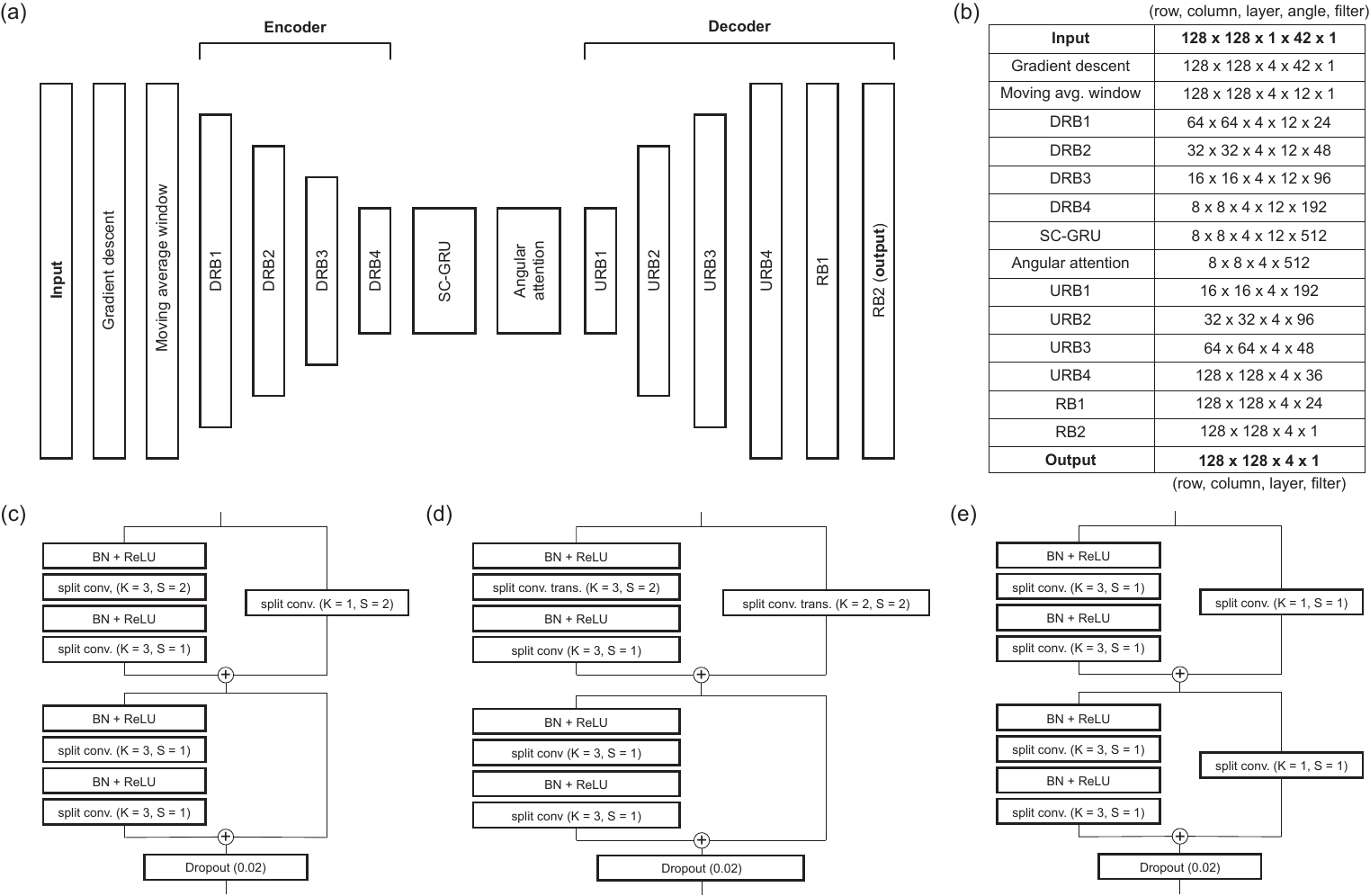}
    \caption{Details on implementing the dynamical scheme of Figure~\protect{\ref{fig:introduction}}. (a) Overall network architecture; (b) tensorial dimensions of each layer; (c) down-residual block (DRB); (d) up-residual block (URB); and (e) residual block (RB). $K$ and $S$ indicate the sizes of kernel and stride, respectively, and the values shown apply only to the row and column axes. For the layer axis, $K=4$ and $S=1$ always. The disparities are to implement the split convolution scheme; please see Section~\protect{\ref{sec:comput_arch}}.\protect{\ref{subsec:sc_gru}} and Figure~\protect{\ref{fig:split_convolution}}.}
    \label{fig:architecture}
\end{figure*}

The proposed RNN architecture is shown in detail in Figure~\ref{fig:architecture}. The forward model and gradient descent Approximant (pre-processing) algorithm are described in Section~\ref{subsec:approximants}. The split-convolutional GRU, convolutional encoder and decoder, and the angular attention mechanism are described in Sections~\ref{subsec:sc_gru}, \ref{subsec:conv_enc_dec}, and \ref{subsec:angular_att}, respectively. The total number of parameters in this computational architecture is $\sim 21\text{M}$ (more on this topic in section~\ref{sec:train_and_test}.\ref{subsec:training_rnn}).

\subsection{Approximant computations}\label{subsec:approximants}

The dense-layered, binary-phase object is illuminated at a sequence of angles, and the corresponding diffraction intensity patterns are captured by a detector. At the $n$-th step of the sequence, the object is illuminated by a plane wave at angles $\left(\theta_{nx},\theta_{ny}\right)$ with respect to the propagation axis $z$ on the $xz$ and $yz$ planes, respectively. Beyond the object, the scattered field propagates in free space by distance $\Delta z$ to the digital camera (the numerical value is $\Delta z=58.2$mm, as we saw in section~\ref{sec:experiment}). Let the forward model under the $n$-th illumination angle be denoted as $H_n$, $n=1,2,\ldots, N$; that is, the $n$-th intensity diffraction pattern at the detector plane produced by the phase object $\mathbf{f}$ is $\mathbf{g}_n\equiv H_n(\mathbf{f})$.

In the simulations, the forward operators $H_n$ are obtained from the non-paraxial beam propagation method (BPM) \cite{feit1980computation,goy2019high,kamilov2016optical}. Let the $j$-th cross-section of the computational window perpendicular to $z$ axis be $f^{[j]} = \exp\left(i\varphi^{[j]}\right),\: j=1,\ldots,J$ where $J$ is the number of slices the we divide the object into, each of axial extent $\delta z$. At the $n$-th illumination angle, the BPM is initialized as $f_n^{[0]}=\text{exp}\left[ik\left(x\sin\theta_{nx}+y\sin\theta_{ny}\right)\right]$, where $k$ is the wavenumber. The optical field at the $(j+1)$-th slice is
\begin{equation}\label{eq:BPM-iteration}
    \begin{split}
        \psi_n^{[j+1]} = \mathcal{F}^{-1}&\bigg[\mathcal{F}\left[\psi_n^{[j]}\circ f_n^{[j]}\right](k_x,k_y)\\
        &\cdot\exp\left(-i\left(k-\sqrt{k^2-k_x^2-k_y^2}\right)\delta z\right)\bigg],
    \end{split}
\end{equation}
where $\delta z$ is equal to the slab thickness, \textit{i.e.} $0.5\:\text{mm}$; ${\cal F}$ and ${\cal F}^{-1}$ are the Fourier and inverse Fourier transforms, respectively; and $\chi_1\circ\chi_2$ denotes the Hadamard (element-wise) product of the functions $\chi_1$, $\chi_2$. The Hadamard product is the numerical implementation of the thin transparency approximation, which is inherent in the BPM. To obtain the intensity at the detector, we define the $(J+1)$-th slice displaced by $\Delta z$ from the $J$-th slice (the latter is the exit surface of the object) to yield 
\begin{equation}\label{eq:forw}
    \mathbf{g}_n\equiv H_n(\mathbf{f})=\left|\psi_n^{[J+1]}\right|^2.
\end{equation}

The purpose of the Approximant, in general, is to produce a crude estimate of the volumetric reconstruction using the forward operator alone. This has been well established as a helpful form of preprocessing for subsequent treatment by machine learning algorithms\ \cite{goy2018low,goy2019high}. Previous works constructed the Approximant as a single-pass gradient descent algorithm \cite{kamilov2016optical,goy2019high}. Here, due to the sequential nature of our reconstruction algorithm, as each intensity diffraction pattern from a new angle of illumination $n$ is received, we instead construct a sequence of Approximants, indexed by $n$, by minimizing the functionals 
\begin{equation}\label{eq:new_loss_function}
    \mathcal{L}_n(\mathbf{f}) = \frac{1}{2}||H_n(\mathbf{f})-\mathbf{g}_n||_2^2,\quad n=1,2,\ldots,N.
\end{equation}
The gradient descent update rule for this functional is
\begin{multline}\label{eq:new_approximants}
    \mathbf{f}_n^{[l+1]} = \mathbf{f}_n^{[l]} -s\left(\nabla_\mathbf{f}\mathcal{L}_n\left(\mathbf{f}_n^{[l]}\right)\right)^\dagger = \\ 
    = \mathbf{f}_n^{[l]} -s\left(H_n^T\left(\mathbf{f}^{[l]}\right)\nabla_\mathbf{f}
    H_n\left(\mathbf{f}_n^{[l]}\right)-\mathbf{g}_n^T\nabla_\mathbf{f}H_n\left(\mathbf{f}_n^{[l]}\right)\right)^\dagger,
\end{multline}
where $\mathbf{f}_n^{[0]}=\mathbf{0}$ and $s$ is the descent step size and in the numerical calculations was set to $0.05$ and the superscript $\dagger$ denotes the transpose. The single-pass, gradient descent-based Approximant was used for training of the RNN but with an additional pre-processing step that will be explained in (\ref{eq:moving_window}).

We also implemented a denoised Total Variation (TV) based Approximant, to be used only at the testing stage of the RNN. In this case, the functional to be minimized is
\begin{equation}\label{eq:TV_Approx}
    \mathcal{L}^{\text{TV}}_n(\mathbf{f}) = \frac{1}{2} ||H_n(\mathbf{f})-\mathbf{g}||_2^2 + \kappa\text{TV}_{l_1}(\mathbf{f}),\quad n=1,2,\ldots,N,
\end{equation}
where the TV-regularization parameter was chosen as $\kappa=10^{-3}$, and for $\mathbf{x}\in \mathcal{R}^{P\times Q}$ the anisotropic $l_1$-TV operator is
\begin{equation}
    \begin{split}
        \text{TV}_{l_1}(\mathbf{x}) = &\sum_{p=1}^{P-1}\sum_{q=1}^{Q-1} \Big(\left|x_{p,q} - x_{p+1,q}\right| + \left|x_{p,q} - x_{p,q+1}\right|\Big)\\
        & + \sum_{p=1}^{P-1} \left|x_{p,Q}-x_{p+1,Q}\right| +\sum_{q=1}^{Q-1} \left|x_{P,q}-x_{P,q+1}\right| 
    \end{split}
\end{equation}
with reflexive boundary conditions \cite{beck2009fast,chambolle2004algorithm}. To produce the Approximants for testing from this functional, we first ran $3$ iterations of the gradient descent and ran $2$ iterations of the FGP-FISTA (Fast Gradient Projection with Fast Iterative Shrinkage Thresholding Algorithm)\ \cite{beck2009fast,beck2009fista}.

The sequence of $N$ Approximants for either training or testing procedure is a $4$D spatiotemporal sequence $\mathbf{F}=\left(\mathbf{f}_1^{[1]},\mathbf{f}_2^{[1]},\ldots,\mathbf{f}_N^{[1]}\right)$. As an additional processing step, to suppress unwanted artifacts in the Approximants of the experimentally captured intensities $\mathbf{g}_n$, we reduce the sequence size to $M$ by applying a moving average window as
\begin{equation}\label{eq:moving_window}
    \tilde{\mathbf{f}}_m^{[1]} = 
    \begin{dcases}
        \frac{1}{N_{\text{w}}+1}\sum_{n=m}^{m+N_{\text{w}}} \mathbf{f}_n^{[1]}, & 1\leq m\leq N_{\text{h}}\\
        \frac{1}{N_{\text{w}}+1}\sum_{n=m}^{m+N_{\text{w}}} \mathbf{f}_{n+N_{\text{w}}}^{[1]}, & N_{\text{h}}+1\leq m\leq M.
    \end{dcases}
\end{equation}
To be consistent, the moving average window was applied to the Approximants for both training and testing. In this study, $N_{\text{w}}=15$, $N_{\text{h}}=6$ and $M=12$. These choices follow from the following considerations. We have $N=42$ diffraction patterns for each sequence: $21$ captured along the $x$ axis ($1-21$) and the remaining ones along the $y$ axis ($22-42$). The window is first applied to $21$ patterns from $x$-axis rotation, which thus generates $6$ averaged diffraction patterns, and then the window is applied to the remaining $21$ patterns from $y$-axis rotation, resulting in the other $6$ patterns. Therefore, the input sequence to the next step in the architecture of Figure~\ref{fig:architecture}, {\it i.e.} to the encoder (Section~\ref{subsec:conv_enc_dec}), consists of a sequence of $M=12$ averaged Approximants~$\tilde{\mathbf{f}}_m^{[1]}$.

\subsection{Split-convolutional gated recurrent unit (SC-GRU)}\label{subsec:sc_gru}

Recurrent neural networks involve a recurrent unit that retains memory and context based on previous inputs in a form of latent tensors or hidden units. It is well known that the Long Short-Term Memory (LSTM) is robust to instabilities in the training process. Moreover, in the LSTM, the weights applied to past inputs are updated according to usefulness, while less useful past inputs are forgotten. This encourages the most salient aspects of the input sequence to influence the output sequence\ \cite{hochreiter1997long}. Recently, the Gated Recurrent Unit (GRU) was proposed as an alternative to LSTM. The GRU effectively reduces the number of parameters by merging some operations inside the LSTM, without compromising quality of reconstructions; thus, it is expected to generalize better in many cases\ \cite{cho2014learning}. For this reason, we chose to utilize the GRU in this paper as well.

The governing equations of the standard GRU are as follows:
\begin{equation}\label{eq:gru_equations}
    \begin{gathered}
        r_m = W_r \xi_m + U_r h_{m-1}+b_r\\
        z_m = W_z \xi_m + U_zh_{m-1} + b_z\\
        \Tilde{h}_m = \text{tanh}\left(W\xi_m+U\left(r_m\circ h_{m-1}\right)+b_h\right)\\
        h_m = (1-z_m)\circ \Tilde{h}_m + z_m\circ h_{m-1},
    \end{gathered}
\end{equation}
where $\xi_m$, $h_m$, $r_m$, $z_m$ are the inputs, hidden features, reset states, and update states, respectively. Multiplication operations with weight matrices are performed in a fully connected fashion. 

We modified this architecture so as to take into account the asymmetry between the lateral and axial dimensions of optical field propagation. This is evident even in free-space propagation, where the lateral components of the Fresnel kernel 
\[
\expb{i\pi\frac{x^2+y^2}{\lambda z}}
\]
are shift invariant and, thus, convolutional, whereas the longitudinal axis $z$ is not. The asymmetry is also evident in nonlinear propagation, as in the BPM forward model (\ref{eq:BPM-iteration}) that we used here. This does not mean that space is anisotropic --- of course space is isotropic! The asymmetry arises because propagation and the object are 3D, whereas the sensor is 2D. In other words, the orientation of the image plane breaks the symmetry in object space so that the scattered field from a certain voxel within the object {\em apparently} influences the scattered intensity from its neighbors at the detector plane differently in the lateral direction than in the axial direction. To account for this asymmetry in a profitable way for our learning task, we first define the operators $W_r$, $U_r$, etc. as convolutional so as to keep the number of parameters down (even though in free space propagation the axial dimension is not convolutional and under strong scattering neither dimension is nonlinear); and we constrain the convolutional kernels of the operators to be the same in the lateral dimensions $x$ and $y$, and allow the axial $z$ dimension kernel to be different. This approach justifies the term Split-Convolutional, and we found it to be a good compromise between facilitating generalization and adhering to the physics of the problem.

\begin{figure}[htbp!]
    \centering
    \includegraphics[width=\linewidth]{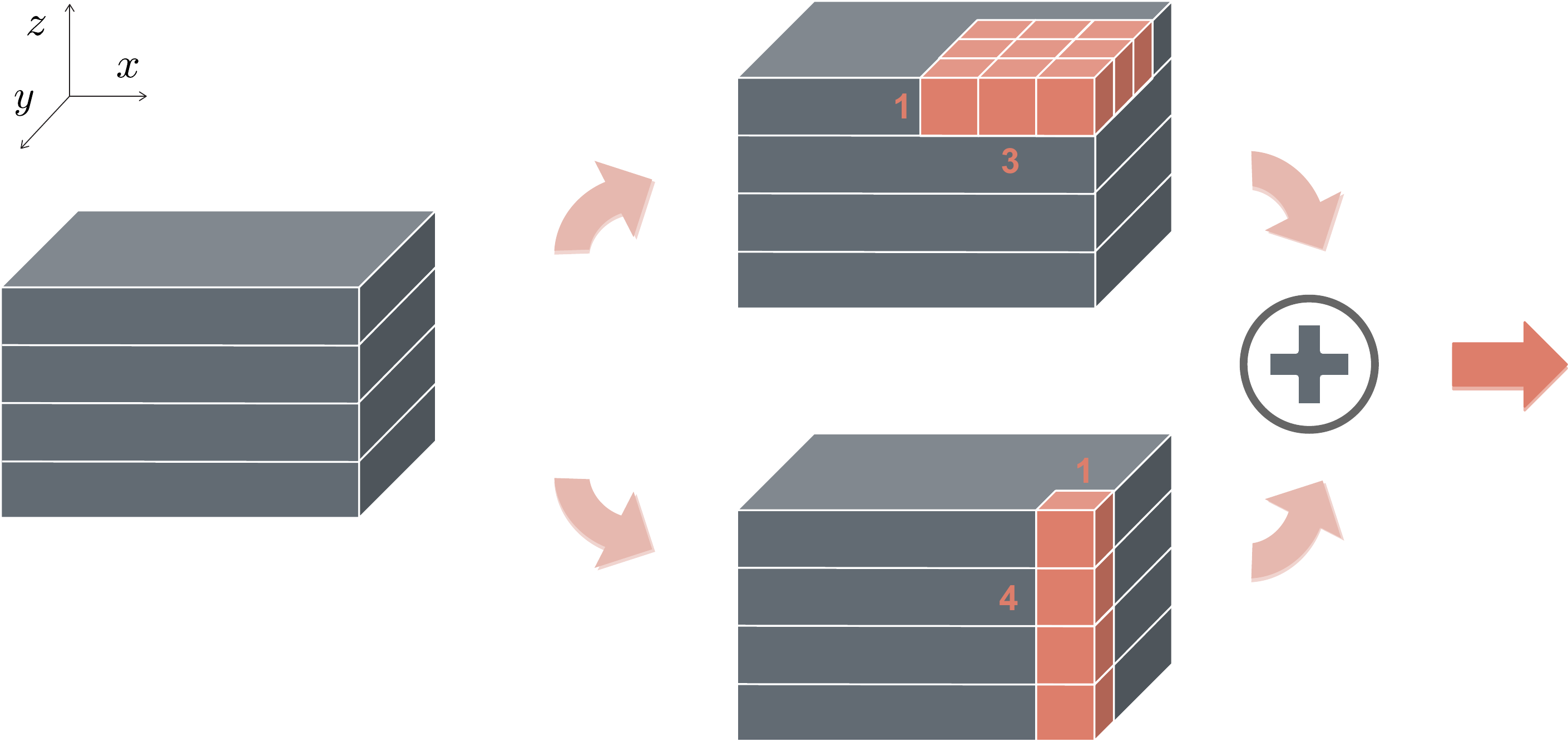}
    \caption{Split convolution scheme: different convolution kernels are applied along the lateral $x,y$ axes {\it vs.} the longitudinal $z$ axis. In our present implementation, the kernels' respective dimensions are $3 \times 3 \times 1$ (or $1 \times 1 \times 1$) and $1 \times 1 \times 4$. The lateral and longitudinal convolutions are computed separately and the results are then added element-wise. The split convolution scheme is used in both the gated recurrent unit (Section~\protect{\ref{sec:comput_arch}}.\protect{\ref{subsec:sc_gru}}) and the encoder/decoder (Section~\protect{\ref{sec:comput_arch}}.\protect{\ref{subsec:conv_enc_dec}}).}
    \label{fig:split_convolution}
\end{figure}

We also replaced the tanh activation function of the standard GRU with a rectified linear unit (ReLU) activation \cite{dey2017gate} as the ReLU is computationally less expensive and helpful to avoid local minima with fewer vanishing gradient problems \cite{nair2010rectified,glorot2011deep}. The final form of our SC-GRU dynamics is
\begin{equation}\label{eq:new_gru_equations}
    \begin{gathered}
        r_m = W_r*\xi_m + U_r*h_{m-1}+b_r\\
        z_m = W_z*\xi_m + U_z*h_{m-1} + b_z\\
        \Tilde{h}_m = \text{ReLU}\left(W*\xi_m+U*\left(r_m\circ h_{m-1}\right)+b_h\right)\\
        h_m = (1-z_m)\circ \Tilde{h}_m + z_m\circ h_{m-1},
    \end{gathered}
\end{equation}
where $*$ denotes our split convolution operation.

\subsection{Convolutional encoder and decoder}\label{subsec:conv_enc_dec}

Convolutional neural networks (CNNs) are placed before and after the SC-GRU as encoder and decoder, respectively. This architectural choice was inspired by \cite{sinha2017lensless,gehring2016convolutional,hori2017advances,zhao2017learning}. The encoder and decoder also utilize split convolution, as shown in Figure~\ref{fig:split_convolution}, in conjunction with residual learning, which is known to improve generalization in deep networks\ \cite{he2016deep}. As in \cite{sinha2017lensless}, the encoder and decoder utilize down-residual blocks (DRB), up-residual blocks (URB), and residual blocks (RB); however, there are no skip connections in our case, {\it i.e.} this is not a U-net\ \cite{ronneberger2015u} architecture. The encoder learns how to map its input ({\it i.e.} the $\tilde{\mathbf{f}}_m^{[1]}$ sequence) onto a low-dimensional nonlinear manifold. The compression factor is $16$ for the lateral input dimensions, but the axial dimension is left intact, as shown in Figure~\ref{fig:architecture}. This eases the burden on the training process as the number of parameters is reduced; more importantly, encoding abstracts features out of the high-dimensional inputs, passing latent tensors over to the recurrent unit. Letting the encoder for the $m$-th angle Approximant be symbolized as $\text{Enc}_m\left(\cdot\right)$, $\xi_m = \text{Enc}_m\left(\tilde{\mathbf{f}}_m^{[1]}\right)$ in (\ref{eq:new_gru_equations}). The decoder restores the output of the RNN to the native dimension of the object we are reconstructing.

\subsection{Angular attention mechanism}\label{subsec:angular_att}

Each intensity diffraction pattern from a new angle of illumination is combined at the SC-GRU input with the hidden feature $h_m$ from the same SC-GRU's previous output. After $M$ iterations, there are $M$ different hidden features resulting from $N$ illumination angles, as seen in (\ref{eq:moving_window}). Since the forward operator $H_n(\mathbf{f})$ is object dependent, the qualitative information that each such new angle conveys will vary with the object. It then becomes interesting to consider whether some angles of illumination convey more information than others.  

The analogue in temporal dynamical systems, the usual domain of application for RNNs, is the {\em attention} mechanism. It decides which elements of the system's state are the most informative. In our case, of course, time has been replaced by the angle of illumination, so we refer to the same mechanism as {\em angular attention:}\ it evaluates the contents of the previously received intensity diffraction patterns from different angles of illumination and assigns to each a compatibility function $e_m$, essentially a weight that is relevant to that illumination's importance for the overall reconstruction. 

Following the summation style attention mechanism\ \cite{bahdanau2014neural}, we compute the compatibility function $e_m$ as output of a neural network with hidden units (layers) $V_e$, $W_e$ and the weights $\alpha_m$ from the compatibility function as 
\begin{equation}\label{eq:attention-VeWe}
    \begin{gathered}
        e_m = V_e\:\text{tanh}\left(W_e h_m\right),  \\
        \alpha_m = \text{softmax}\left(e_m\right) = \frac{\text{exp}(e_m)}{\sum_{m=1}^{M} \text{exp}(e_m)}, \\
        \quad m = 1,2,\ldots, M.
    \end{gathered}
\end{equation}
The final angular attention output $a$ is then computed from a linear combination of the hidden features as
\begin{equation}\label{eq:attention}
        a=\sum_{m=1}^{M} \alpha_m h_m.
\end{equation}
For the ablation study of Section~\ref{sec:results}, only the last hidden feature $h_M$ is passed on to the decoder, {\it i.e.} the angular attention mechanism is not used. There is an alternative, dot-product attention mechanism\ \cite{vaswani2017attention}, but we chose not to implement it here.

\section{Training and testing procedures}\label{sec:train_and_test}

\subsection{Training the recurrent neural network}\label{subsec:training_rnn}

For training and validation, $5000$ and $500$ layered objects were used, respectively. For each object, a sequence of intensity diffraction patterns from the $N=42$ angles of illumination was produced by BPM, as described earlier. The Approximants were obtained each as a single iteration of the gradient descent. All of the architectures were trained for $100$ epochs with a training loss function (TLF) of negative Pearson correlation coefficient (NPCC) \cite{li2018imaging}, defined as
\begin{equation}
    \sst{\mathcal{E}}{NPCC}\big(f,\hat{f}\big) \equiv  -\:\frac{\displaystyle{\sum_{x,y}}\Big(f(x,y)-\big<f\big>\Big)\Big(\hat{f}(x,y)-\big<\hat{f}\big>\Big)}{\sqrt{\displaystyle{\sum_{x,y}}\Big(f(x,y)-\big<f\big>\Big)^2}\sqrt{\displaystyle{\sum_{x,y}}\Big(\hat{f}(x,y)-\big<\hat{f}\big>\Big)^2}}, \label{eq:tlf-npcc}
\end{equation}
where $f$ and $\hat{f}$ are a ground truth image and its corresponding reconstruction. In this article, our NPCC function was defined to perform computation in $3$D. We used a stochastic gradient descent scheme with the \textit{Adam} optimizer \cite{kingma2014adam}. The learning rate was set to be $10^{-3}$ initially and halved whenever validation loss plateaued for $5$ consecutive epochs. Batch size was set to be $10$. The desktop computer used for training has Intel Xeon W-$2295$ CPU at $3.00$ GHz with $24.75$ MB cache, $128$ GB RAM, and dual NVIDIA Quadro RTX $8000$ GPUs with $48$ GB VRAM.

For comparison, we also re-trained the $3$D-DenseNet architecture with skip connections in \cite{goy2019high} with the same training scheme above, \textit{i.e.}  on \textit{Adam} for $100$ epochs and with batch size of $10$ and the same learning rate initial value and halving strategy. This serves as baseline; however, the number of parameters in this network is $0.5\:\text{M}$, whereas in our RNN architecture the number of parameters is $21\:\text{M}$. We also trained an enhanced version of the $3$D-DenseNet by tuning the number of dense blocks, the number of layers inside each dense block, filter size, and growth rate to match the total number of parameters with that of the RNN, {\it i.e.} $21\:\text{M}$. In the next section, we refer to these two versions of the $3$D-DenseNet as Baseline ($0.5\:\text{M}$) and Baseline ($21\:\text{M}$), respectively.

\subsection{Testing procedures and metrics}

A simple affine transform is first applied to the raw experimentally obtained intensity diffraction patterns to correct slight misalignment. Then we run the gradient descent method up to $3$ iterations of the gradient descent (\ref{eq:new_approximants}) and the FGP-FISTA up to $2$ iterations, to test the trained network using the TV-based Approximants (\ref{eq:TV_Approx}).

Even though training used NPCC as in (\ref{eq:tlf-npcc}), we investigated two additional metrics for testing: probability of error (PE), the Wasserstein distance \cite{villani2003topics,kolouri2017optimal}. We also quantified test performance using the SSIM (Structural Similarity Index Metric) \cite{wang2004image}, shown in the Supplementary material. 

PE is the mean absolute error between two binary objects; in the digital communication community it is instead referred to as Bit Error Rate (BER). To obtain the PE, we first threshold the reconstructions and then define
\begin{equation}
    \text{PE} = \frac{\left(\text{\# false negatives}\right)\: + \:\left(\text{\# false positives}\right)}{\text{total \# pixels}}.
\end{equation}
We found that it oftentimes helps to accentuate the differences between a binary phase ground truth object and its binarized reconstruction as even small residual artifacts, if they are above the threshold, are thresholded to be one, and thus they are taken into account to the probability of error calculation more than they would have been to other metrics. With these procedures, PE is a particularly suitable error metric for the kind of objects we consider in this paper.

PE is also closely related to the two-dimensional Wasserstein distance as we will now show through an analytical derivation. The latter metric involves an optimization process in terms of a transport plan to minimize the total cost of transport from a source distribution to a target distribution. The two-dimensional Wasserstein distance is defined as
\begin{equation}
    \begin{gathered}
        W_{p=1} = \min_P \langle P,C\rangle = \min_P\sum_{ij}\sum_{kl}\gamma_{ij,kl}C_{ij,kl},\\
        \text{s.t.}\:\: \sum_{kl}\gamma_{ij,kl} = f_{ij},\: \sum_{ij}\gamma_{ij,kl}=g_{kl},\:\gamma_{ij,kl}\geq 0,
    \end{gathered}
\end{equation}
where $f_{ij}$ and $g_{kl}$ are a ground truth binary object and its binary reconstruction, \textit{i.e.} $f_{ij}, g_{kl}, \gamma_{ij,kl} \in\{0,1\}$, a coupling tensor $P=\left(\gamma_{ij,kl}\right)$, and a cost tensor $C_{ij,kl}=\left|x_{ij}-x_{kl}\right|$. PE can be reduced to have a similar, but not equivalent, form to that of the Wasserstein distance. For $i,j,k,l$ where $\gamma_{ij,kl}\neq 0$,
\begin{equation}\label{eq:prob}
    \begin{split}
        \text{PE} &= \frac{1}{N^2}\sum_{ij}\left|f_{ij}-g_{ij}\right|\\
        &= \frac{1}{N^2}\sum_{ij}\left|f_{ij}-\sum_{kl}g_{kl}\:\delta\left[i-k,j-l\right]\right|\\
        &= \frac{1}{N^2}\sum_{ij}\left|\sum_{kl}\gamma_{ij,kl}\left(1-\frac{g_{kl}\:\delta\left[i-k,j-l\right]}{\gamma_{ij,kl}}\right)\right|\\
        &\equiv \sum_{ij}\left|\sum_{kl}\gamma_{ij,kl}\Tilde{C}_{ij,kl}\right|\\
        &= \sum_{ij,kl;\gamma_{ij,kl}\neq 0} \gamma_{ij,kl}\Tilde{C}_{ij,kl}, \qquad \text{where}
    \end{split}
\end{equation}
\begin{equation}
        N^2\Tilde{C}_{ij,kl} = 1\!-\!\frac{g_{kl}\:\delta\left[i-k,j-l\right]}{\gamma_{ij,kl}}=
        \begin{dcases}
            \:1, & \:\text{if}\:\: ij \neq kl\\
            \:1\! -\! g_{kl}, & \:\text{if}\:\: ij = kl.
        \end{dcases}
\end{equation}
This shows that the PE is a version of the Wasserstein distance with differently defined cost tensor.

\section{Results}\label{sec:results}


\begin{figure*}[htbp]
    \centering
    \includegraphics[width=0.55\textwidth]{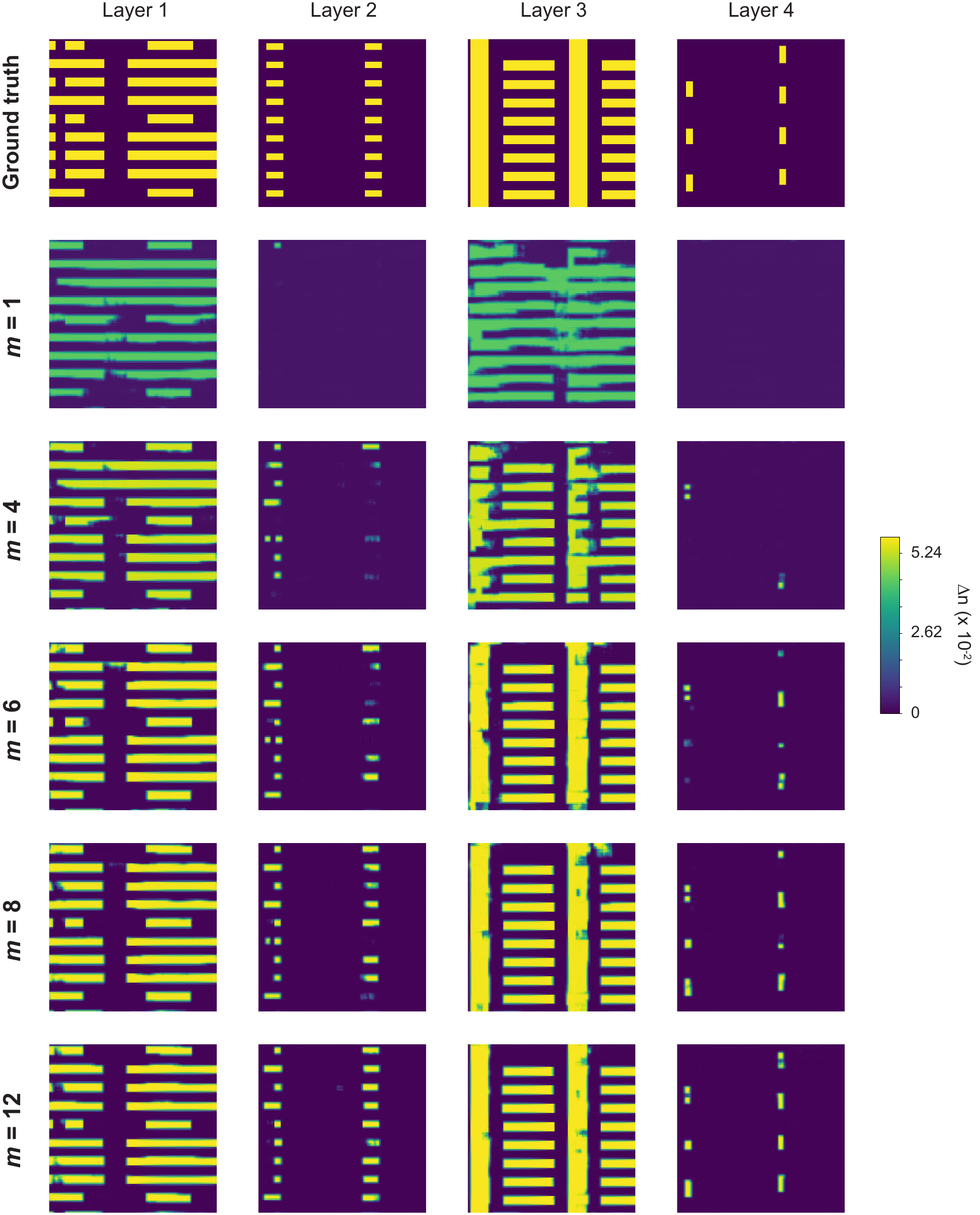}
    \caption{Progress of 3D reconstruction performance as new windowed Approximants $m=1,\ldots, M\!\!=\!\!12$ according to (\protect{\ref{eq:moving_window}}) applied on experimental data are presented to the recurrent scheme. The same progression can be found in the Online Materials as a movie.}
    \label{fig:number_of_patterns}
\end{figure*}

\begin{table*}[htbp!]
    \begin{center}
    \begin{tabular}{c||c c c c|c}
        \hline
        \textbf{Probability of error ($\%$)} ($\downarrow$) & \text{Layer} 1 & \text{Layer} 2 & \text{Layer} 3 & \text{Layer} 4 & \text{Overall}\\
        \hline
        \text{Baseline (0.5 M)} & 6.604 & 5.255 & 7.837 & 3.204 & 5.725\\
        \text{Baseline (21 M)} & 6.604 & 5.725 & 5.652 & 2.856 & 5.209\\
        \hdashline
        \text{Proposed RNN (21 M)} & \textbf{5.408} & \textbf{4.828} & \textbf{2.332} & \textbf{1.660} & \textbf{3.557}\\
        \hline\hline
        \textbf{Wasserstein distance} ($\times\:10^{-2}$) ($\downarrow$) & \text{Layer} 1 & \text{Layer} 2 & \text{Layer} 3 & \text{Layer} 4 & \text{Overall}\\
        \hline
        \text{Baseline (0.5 M)} & 2.854 & 1.466 & 2.783 & 0.9900 & 2.023\\
        \text{Baseline (21 M)} & 2.703 & 1.171 & 2.475 & 0.8112 & 1.790\\
        \hdashline
        \text{Proposed RNN (21 M)} & \textbf{1.999} & \textbf{1.093} & \textbf{1.749} & \textbf{0.6403} & \textbf{1.370}\\
        \hline\hline
        \textbf{PCC} ($\uparrow$) & \text{Layer} 1 & \text{Layer} 2 & \text{Layer} 3 & \text{Layer} 4 & \text{Overall}\\
         \hline
        \text{Baseline (0.5 M)} & 0.8818 & 0.6426 & 0.8658 & 0.6191 & 0.7523\\
        \text{Baseline (21 M)} & 0.8859 & 0.6430 & 0.9021 & 0.6132 & 0.7611\\
        \hdashline
        \text{Proposed RNN (21 M)} & \textbf{0.8943} & \textbf{0.6612} & \textbf{0.9551} & \textbf{0.7039} & \textbf{0.8036}\\
        \hline
    \end{tabular}
    \end{center}
    \caption{Quantitative comparison between the baseline (static) and dynamic reconstruction from testing on experimental data, according to PE, Wasserstein distance ($p=1$), and PCC. SSIM comparisons are in the Supplementary materials.}
    \label{tab:quantitative_comparison}
\end{table*}

Our RNN is first trained as described in Section~\ref{sec:train_and_test}, and then tested with the TV-based Approximants (\ref{eq:TV_Approx}) applied to the experimentally obtained diffraction patterns. The evolution of the RNN output as more input patterns are presented is shown in Figure~\ref{fig:number_of_patterns}. When the recurrence starts with $m=1$, the volumetric reconstruction is quite poor; as more orientations are included, the reconstruction improves as expected. A movie version of this evolution for $m=1,\ldots, M$ is included in the online materials. 


\begin{figure*}[htbp!]
    \centering
    \includegraphics[width=0.72\textwidth]{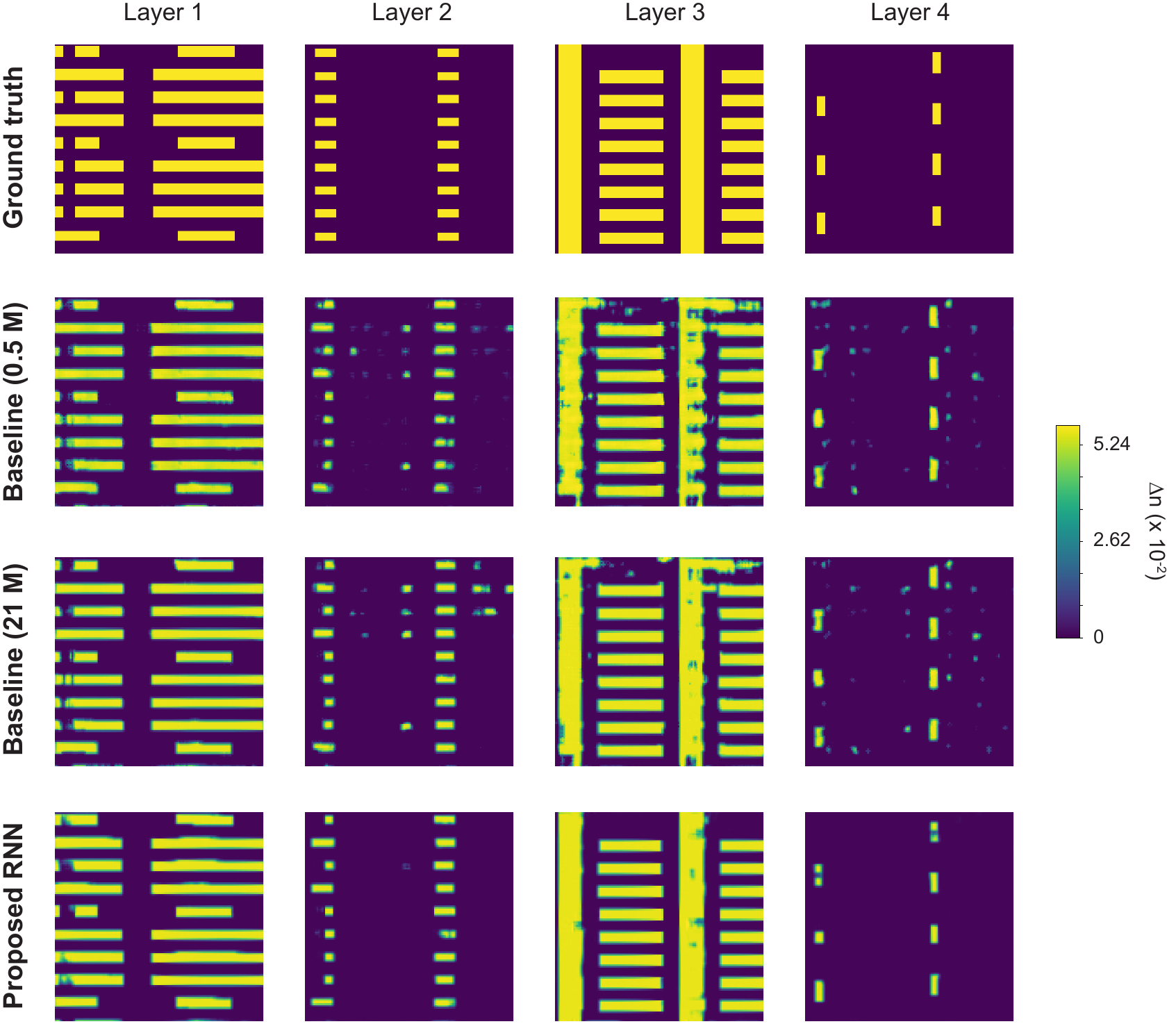}
    \caption{Qualitative comparison on test performance between the baseline and proposed architectures using experimental data. The baseline architectures are $3$D-DenseNet CNN architectures with $0.5$ M and $21$ M parameters. Our proposed architecture is a recurrent neural network with elements described in Section~\ref{sec:comput_arch}.}
    \label{fig:qualitative_comparison}
\end{figure*}

\begin{table*}[htbp!]
    \begin{center}
    \begin{tabular}{c||c c c c|c}
        \hline
        \textbf{Probability of error ($\%$)} ($\downarrow$) & \text{Layer} 1 & \text{Layer} 2 & \text{Layer} 3 & \text{Layer} 4 & \text{Overall}\\
        \hline
        \text{Proposed RNN (21 M)} & \textbf{5.408} & 4.828 & \textbf{2.332} & \textbf{1.660} & \textbf{3.557}\\
        \hdashline
        \text{-- ReLU activation (21 M)} & 6.262 & \textbf{4.718} & 3.241 & 1.904 & 4.031\\
        \text{-- angular attention (21 M)} & 9.399 & 5.566 & 11.64 & 3.375 & 7.495\\
        \text{-- split convolution (43 M)} & 9.674 & 6.342 & 14.43 & 2.405 & 8.212\\
        \hline\hline
        \textbf{Wasserstein distance} ($\times\:10^{-2}$) ($\downarrow$) & \text{Layer} 1 & \text{Layer} 2 & \text{Layer} 3 & \text{Layer} 4 & \text{Overall}\\
        \hline
        \text{Proposed RNN (21 M)} & \textbf{1.999} & \textbf{1.093} & \textbf{1.749} & \textbf{0.6403} & \textbf{1.370}\\
        \hdashline
        \text{-- ReLU activation (21 M)} & 2.291 & 1.156 & 1.886 & 0.6692 & 1.501\\
        \text{-- angular attention (21 M)} & 3.016 & 1.587 & 3.672 & 1.063 & 2.335\\
        \text{-- split convolution (43 M)} & 4.005 & 2.863 & 3.651 & 2.233 & 3.188\\
        \hline\hline
        \textbf{PCC} ($\uparrow$) & \text{Layer} 1 & \text{Layer} 2 & \text{Layer} 3 & \text{Layer} 4 & \text{Overall}\\
        \hline
        \text{Proposed RNN (21 M)} & \textbf{0.8943} & 0.6612 & \textbf{0.9551} & \textbf{0.7039} & \textbf{0.8036}\\
        \hdashline
        \text{-- ReLU activation (21 M)} & 0.8832 & \textbf{0.6836} & 0.9406 & 0.6725 & 0.7950\\
        \text{-- angular attention (21 M)} & 0.8281 & 0.6252 & 0.8145 & 0.4657 & 0.6834\\
        \text{-- split convolution (43 M)} & 0.8005 & 0.4525 & 0.7313 & 0.4910 & 0.6188\\
        \hline
    \end{tabular}
    \end{center}
    \caption{Quantitative assessment of ablation effects. Values inside the parentheses in the first column indicate the number of parameters. When we ablate split convolution, we rather choose $3\times 3\times 3$ being the uniform kernel, and, hence, the number of parameters increases. SSIM comparisons are in the Supplementary materials.}
    \label{tab:ablation_study_quantitative}
\end{table*}

Visual comparisons with the baseline $3$D-DenseNets with $0.5$ M and $21$ M parameters are shown in Figure~\ref{fig:qualitative_comparison}. The RNN results show substantial visual improvement, with fewer artifacts and distortions compared to static approaches, e.g. \cite{goy2019high}. Quantitatively comparisons in terms of our chosen metrics PE, Wasserstein, and PCC are in Table~\ref{tab:quantitative_comparison}.


\begin{figure*}[t!]
    \centering
    \includegraphics[width=0.61\textwidth]{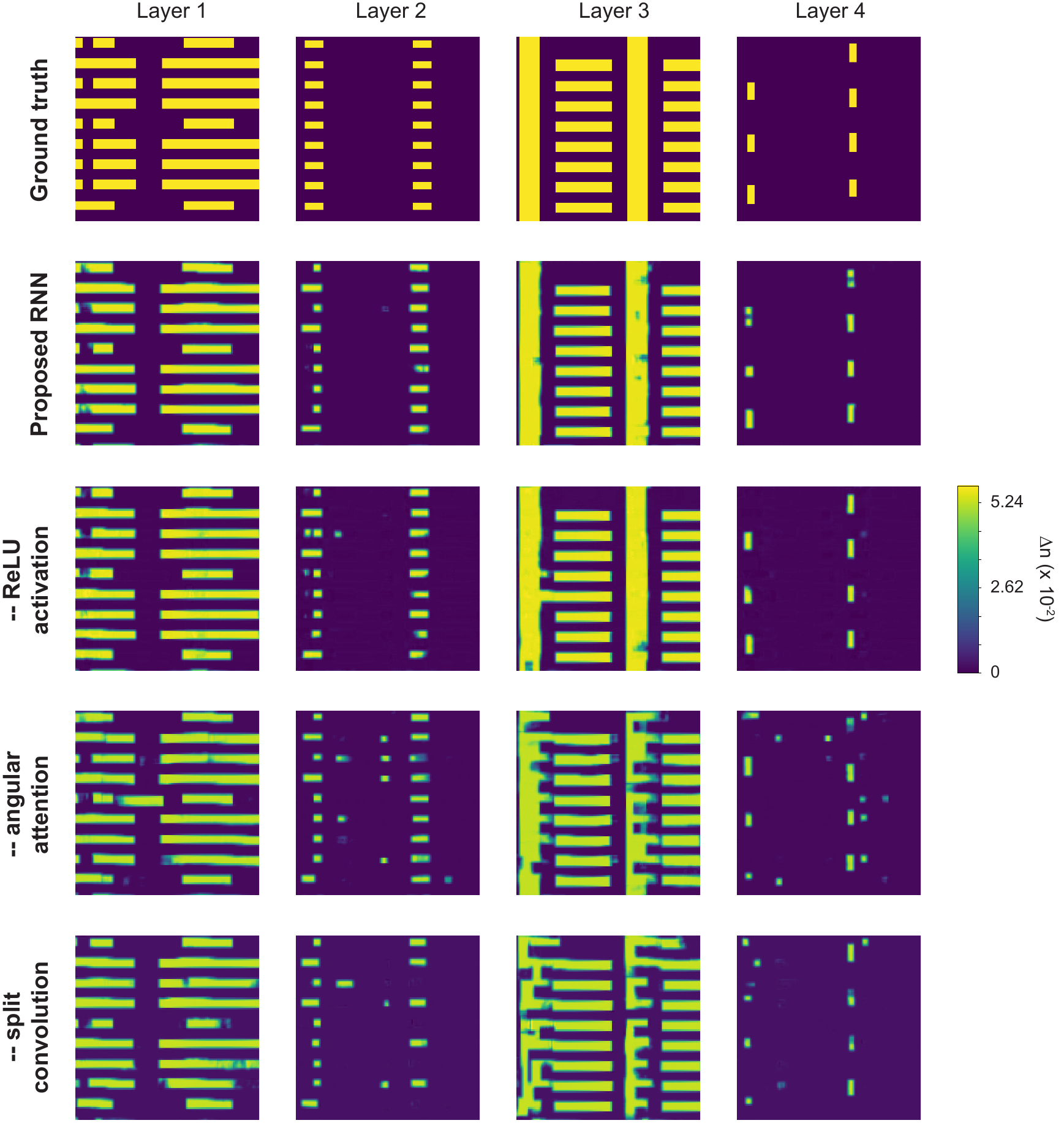}
    \caption{Visual quality assessment from the ablation study on elements described in Section~\ref{sec:comput_arch}. Rows $3-5$ show reconstructions based on experimental data for each layer upon ablation of ReLU activation (\ref{eq:new_gru_equations}), {\it i.e.}, using the more common tanh activation function instead (row 3); angular attention mechanism (row 4); and split convolution (row 5). The rows are ordered by increasing severity of the ablation effect.}
    \label{fig:ablation_study_qualitative}
\end{figure*}

We conducted an ablation study, and its purpose is to isolate and compare quantitatively the contribution to the reconstruction of each element described in Figure~\ref{fig:architecture} and Section~\ref{sec:comput_arch}. We remove, one at a time, the split convolution, angular attention mechanism, and ReLU activation, and quantify performance again. Ablation in the case of ReLU activation means that we replace it with the tanh activation function, which is more usual. The ablated architectures are also trained under the same training scheme in Section~\ref{sec:train_and_test}.\ref{subsec:training_rnn} and tested with the same TV-based Approximants. 

Visually, the ablation of the split convolution affects and degrades the testing performance worst, followed by the ablation of the angular attention mechanism and the ReLU activation. These findings are supported quantitatively as well in Table~\ref{tab:ablation_study_quantitative}. Note that the substitution of the ReLU with the tanh does not bring a large increase compared to others, but even slightly better in some case (see the probability of error of Layer $2$ in Table~\ref{tab:ablation_study_quantitative}). Thus, we find that (1) the split convolution should be considered to replace a general $3$D convolution when designing a recurrent unit and a convolutional encoder/decoder; (2) the angular attention mechanism is helpful when the inputs are formulated into temporal sequences; and (3) the choice of ReLU over tanh is still helpful but somewhat less significant and may be application-dependent. With respect to attention, in particular, even though the module's presence clearly contributes to good training quality, we found that the coefficients converge to $\alpha_m\approx 1/M$ for all $m$, consistent with the more-or-less angularly invariant class of samples---at least in the statistical sense, and for the small range of illumination angles that we used. A more detailed study of the angular attention module can be found in the Supplementary Material.

\section{Conclusions and discussion}\label{sec:conclusion}

We have proposed a radically new recurrent neural network scheme for processing raw inputs from different angles of illumination dynamically, {\it i.e.} as a sequence, with each new angle improving the 3D reconstruction. We have found this scheme to offer significant qualitative and quantitative improvement over static machine learning schemes, where the raw inputs from all angles are processed at once by a neural network. Through an ablation study, we found that sandwiching the recurrent structure between a convolutional encoder/decoder helps improve the reconstructions. Even more interestingly, an angular attention mechanism, rewarding raw inputs from certain angles as more informative and penalizing others, also contributes significantly to improving reconstruction fidelity albeit less than the encoder/decoder pair. 

Even though we used the dynamic machine learning approach in the most difficult case of 3D reconstruction when strong scattering is present, there is no reason to doubt that it would be applicable to less ill-posed cases as well, e.g. optical diffraction tomography and Radon inverse. Also possible are alternative implementations of the RNN, e.g. with LSTMs or Reservoir Computing \cite{lukovsevivcius2009reservoir,lukovsevivcius2012reservoir,schrauwen2007overview}, and further exploration of split convolutional variants or DenseNet variants for the encoder/decoder and dynamical units; we leave these investigations to future work. 

\section{Funding}
\noindent Southern University of Science and Technology (6941806); Intelligence Advanced Research Projects Activity (FA8650-17-C-9113); Korea Foundation for Advanced Studies.~\\

\section{Acknowledgments}
\noindent I. Kang acknowledges partial support from KFAS (Korea Foundation for Advanced Studies) scholarship. We are grateful to Jungmoon Ham for her assistance with drawing Figures~\ref{fig:introduction} and \ref{fig:split_convolution}, and to Subeen Pang, Mo Deng and Peter So for useful discussions and suggestions.~\\

\noindent\textbf{Disclosures.} The authors declare no conflicts of interest.

\bibliography{recurrent,tomography,deep_learning,algorithm,holography,inverse}

\begin{thebibliography}{10}
\newcommand{\enquote}[1]{``#1''}

\bibitem{radon1986determination}
J.~Radon, \enquote{On the determination of functions from their integral values
  along certain manifolds,} {\protect\JournalTitle{IEEE Trans. Med. Imaging}}
  \textbf{5}, 170--176 (1986). Translated by P. C. Parks from the original
  German text.

\bibitem{radon1917determination}
J.~Radon, \enquote{On the determination of functions from their integrals along
  certain manifolds,} {\protect\JournalTitle{Ber. Verh, Sachs Akad Wiss.}}
  \textbf{69}, 262--277 (1917).

\bibitem{bracewell1967inversion}
R.~N. Bracewell and A.~Riddle, \enquote{Inversion of fan-beam scans in radio
  astronomy,} {\protect\JournalTitle{Astrophys. J.}} \textbf{150}, 427 (1967).

\bibitem{feldkamp1984practical}
L.~A. Feldkamp, L.~C. Davis, and J.~W. Kress, \enquote{Practical cone-beam
  algorithm,} {\protect\JournalTitle{J. Opt. Soc. Am. A}} \textbf{1}, 612--619
  (1984).

\bibitem{dreike1976convolution}
P.~Dreike and D.~P. Boyd, \enquote{Convolution reconstruction of fan beam
  projections,} {\protect\JournalTitle{Comp. Graph. Image Process.}}
  \textbf{5}, 459--469 (1976).

\bibitem{wang1993general}
G.~Wang, T.-H. Lin, P.-c. Cheng, and D.~M. Shinozaki, \enquote{A general
  cone-beam reconstruction algorithm,} {\protect\JournalTitle{IEEE Trans. Med.
  Imaging}} \textbf{12}, 486--496 (1993).

\bibitem{kudo1991helical}
H.~Kudo and T.~Saito, \enquote{Helical-scan computed tomography using cone-beam
  projections,} in \emph{Conference Record of the 1991 IEEE Nuclear Science
  Symposium and Medical Imaging Conference,}  (IEEE, 1991), pp. 1958--1962.

\bibitem{grangeat1991mathematical}
P.~Grangeat, \enquote{Mathematical framework of cone beam {3D} reconstruction
  via the first derivative of the {Radon} transform,} in \emph{Mathematical
  Methods in Tomography,}  (Springer, 1991), pp. 66--97.

\bibitem{katsevich2002analysis}
A.~Katsevich, \enquote{Analysis of an exact inversion algorithm for spiral
  cone-beam {CT},} {\protect\JournalTitle{Phys. Med. Biol.}} \textbf{47}, 2583
  (2002).

\bibitem{choi2007tomographic}
W.~Choi, C.~Fang-Yen, K.~Badizadegan, S.~Oh, N.~Lue, R.~R. Dasari, and M.~S.
  Feld, \enquote{Tomographic phase microscopy,} {\protect\JournalTitle{Nat.
  Methods}} \textbf{4}, 717--719 (2007).

\bibitem{ishimaru2017electromagnetic}
A.~Ishimaru, \emph{Electromagnetic wave propagation, radiation, and scattering:
  from fundamentals to applications} (John Wiley \& Sons, 2017).

\bibitem{tatarski2016wave}
V.~I. Tatarski, \emph{Wave propagation in a turbulent medium} (Courier Dover
  Publications, 2016).

\bibitem{wolf1969three}
E.~Wolf, \enquote{Three-dimensional structure determination of semi-transparent
  objects from holographic data,} {\protect\JournalTitle{Opt. Commun.}}
  \textbf{1}, 153--156 (1969).

\bibitem{devaney1981inverse}
A.~Devaney, \enquote{Inverse-scattering theory within the {Rytov}
  approximation,} {\protect\JournalTitle{Opt. Lett.}} \textbf{6}, 374--376
  (1981).

\bibitem{pham2020three}
T.-a. Pham, E.~Soubies, A.~Ayoub, J.~Lim, D.~Psaltis, and M.~Unser,
  \enquote{Three-dimensional optical diffraction tomography with
  {Lippmann-Schwinger} model,} {\protect\JournalTitle{IEEE Trans. Comput.
  Imaging}} \textbf{6}, 727--738 (2020).

\bibitem{marks2006family}
D.~L. Marks, \enquote{A family of approximations spanning the {Born} and
  {Rytov} scattering series,} {\protect\JournalTitle{Opt. Express}}
  \textbf{14}, 8837--8848 (2006).

\bibitem{milgram2002computational}
J.~H. Milgram and W.~Li, \enquote{Computational reconstruction of images from
  holograms,} {\protect\JournalTitle{Appl. Opt.}} \textbf{41}, 853--864 (2002).

\bibitem{tian2010quantitative}
L.~Tian, N.~Loomis, J.~A. Dom{\'\i}nguez-Caballero, and G.~Barbastathis,
  \enquote{Quantitative measurement of size and three-dimensional position of
  fast-moving bubbles in air-water mixture flows using digital holography,}
  {\protect\JournalTitle{Appl. Opt.}} \textbf{49}, 1549--1554 (2010).

\bibitem{hahn2008wide}
J.~Hahn, H.~Kim, Y.~Lim, G.~Park, and B.~Lee, \enquote{Wide viewing angle
  dynamic holographic stereogram with a curved array of spatial light
  modulators,} {\protect\JournalTitle{Opt. Express}} \textbf{16}, 12372--12386
  (2008).

\bibitem{park2009recent}
J.-H. Park, K.~Hong, and B.~Lee, \enquote{Recent progress in three-dimensional
  information processing based on integral imaging,}
  {\protect\JournalTitle{Appl. Opt.}} \textbf{48}, H77--H94 (2009).

\bibitem{nehmetallah2012applications}
G.~Nehmetallah and P.~P. Banerjee, \enquote{Applications of digital and analog
  holography in three-dimensional imaging,} {\protect\JournalTitle{Adv. Opt.
  Photonics}} \textbf{4}, 472--553 (2012).

\bibitem{williams2013digital}
L.~Williams, G.~Nehmetallah, and P.~P. Banerjee, \enquote{Digital tomographic
  compressive holographic reconstruction of three-dimensional objects in
  transmissive and reflective geometries,} {\protect\JournalTitle{Appl. Opt.}}
  \textbf{52}, 1702--1710 (2013).

\bibitem{brady2009compressive}
D.~J. Brady, K.~Choi, D.~L. Marks, R.~Horisaki, and S.~Lim,
  \enquote{Compressive holography,} {\protect\JournalTitle{Opt. Express}}
  \textbf{17}, 13040--13049 (2009).

\bibitem{choi2010compressive}
K.~Choi, R.~Horisaki, J.~Hahn, S.~Lim, D.~L. Marks, T.~J. Schulz, and D.~J.
  Brady, \enquote{Compressive holography of diffuse objects,}
  {\protect\JournalTitle{Appl. Opt.}} \textbf{49}, H1--H10 (2010).

\bibitem{rivenson2018phase}
Y.~Rivenson, Y.~Zhang, H.~G{\"u}nayd{\i}n, D.~Teng, and A.~Ozcan,
  \enquote{Phase recovery and holographic image reconstruction using deep
  learning in neural networks,} {\protect\JournalTitle{Light Sci. Appl.}}
  \textbf{7}, 17141--17141 (2018).

\bibitem{wu2019bright}
Y.~Wu, Y.~Luo, G.~Chaudhari, Y.~Rivenson, A.~Calis, K.~De~Haan, and A.~Ozcan,
  \enquote{Bright-field holography: cross-modality deep learning enables
  snapshot 3d imaging with bright-field contrast using a single hologram,}
  {\protect\JournalTitle{Light Sci. Appl.}} \textbf{8}, 1--7 (2019).

\bibitem{rivenson2019deep}
Y.~Rivenson, Y.~Wu, and A.~Ozcan, \enquote{Deep learning in holography and
  coherent imaging,} {\protect\JournalTitle{Light Sci. Appl.}} \textbf{8}, 1--8
  (2019).

\bibitem{zhang2018twin}
W.~Zhang, L.~Cao, D.~J. Brady, H.~Zhang, J.~Cang, H.~Zhang, and G.~Jin,
  \enquote{Twin-image-free holography: a compressive sensing approach,}
  {\protect\JournalTitle{Phys. Rev. Lett.}} \textbf{121}, 093902 (2018).

\bibitem{kamilov2016recursive}
U.~S. Kamilov, D.~Liu, H.~Mansour, and P.~T. Boufounos, \enquote{A recursive
  born approach to nonlinear inverse scattering,} {\protect\JournalTitle{IEEE
  Signal Process. Lett.}} \textbf{23}, 1052--1056 (2016).

\bibitem{kamilov2016optical}
U.~S. Kamilov, I.~N. Papadopoulos, M.~H. Shoreh, A.~Goy, C.~Vonesch, M.~Unser,
  and D.~Psaltis, \enquote{Optical tomographic image reconstruction based on
  beam propagation and sparse regularization,} {\protect\JournalTitle{IEEE
  Trans. Comput. Imaging}} \textbf{2}, 59--70 (2016).

\bibitem{giorgi2013application}
G.~Giorgi, M.~Brignone, R.~Aramini, and M.~Piana, \enquote{Application of the
  inhomogeneous {Lippmann--Schwinger} equation to inverse scattering problems,}
  {\protect\JournalTitle{SIAM J. Appl. Math.}} \textbf{73}, 212--231 (2013).

\bibitem{chew1990reconstruction}
W.~C. Chew and Y.-M. Wang, \enquote{Reconstruction of two-dimensional
  permittivity distribution using the distorted {Born} iterative method,}
  {\protect\JournalTitle{IEEE Trans. Med. Imaging}} \textbf{9}, 218--225
  (1990).

\bibitem{sun2018efficient}
Y.~Sun, Z.~Xia, and U.~S. Kamilov, \enquote{Efficient and accurate inversion of
  multiple scattering with deep learning,} {\protect\JournalTitle{Opt.
  Express}} \textbf{26}, 14678--14688 (2018).

\bibitem{lu1985multidimensional}
Z.-Q. Lu, \enquote{Multidimensional structure diffraction tomography for
  varying object orientation through generalised scattered waves,}
  {\protect\JournalTitle{Inverse Probl.}} \textbf{1}, 339 (1985).

\bibitem{lu1986jkm}
Z.-Q. Lu, \enquote{{JKM} perturbation theory, relaxation perturbation theory,
  and their applications to inverse scattering: theory and reconstruction
  algorithms,} {\protect\JournalTitle{IEEE Trans. Ultrason. Ferroelectr. Freq.
  Control}} \textbf{33}, 722--730 (1986).

\bibitem{tsihrintzis2000higher}
G.~A. Tsihrintzis and A.~J. Devaney, \enquote{Higher order (nonlinear)
  diffraction tomography: Inversion of the {Rytov} series,}
  {\protect\JournalTitle{IEEE Trans. Inf. Theory}} \textbf{46}, 1748--1761
  (2000).

\bibitem{feit1980computation}
M.~Feit and J.~Fleck, \enquote{Computation of mode properties in optical fiber
  waveguides by a propagating beam method,} {\protect\JournalTitle{Appl. Opt.}}
  \textbf{19}, 1154--1164 (1980).

\bibitem{kamilov2015learning}
U.~S. Kamilov, I.~N. Papadopoulos, M.~H. Shoreh, A.~Goy, C.~Vonesch, M.~Unser,
  and D.~Psaltis, \enquote{Learning approach to optical tomography,}
  {\protect\JournalTitle{Optica}} \textbf{2}, 517--522 (2015).

\bibitem{shoreh2017optical}
M.~H. Shoreh, A.~Goy, J.~Lim, U.~Kamilov, M.~Unser, and D.~Psaltis,
  \enquote{Optical tomography based on a nonlinear model that handles multiple
  scattering,} in \emph{2017 IEEE International Conference on Acoustics, Speech
  and Signal Processing (ICASSP),}  (Ieee, 2017), pp. 6220--6224.

\bibitem{lim2018learning}
J.~Lim, A.~Goy, M.~H. Shoreh, M.~Unser, and D.~Psaltis, \enquote{Learning
  tomography assessed using {Mie} theory,} {\protect\JournalTitle{Phys. Rev.
  Appl.}} \textbf{9}, 034027 (2018).

\bibitem{chowdhury2019high}
S.~Chowdhury, M.~Chen, R.~Eckert, D.~Ren, F.~Wu, N.~Repina, and L.~Waller,
  \enquote{High-resolution {3D} refractive index microscopy of
  multiple-scattering samples from intensity images,}
  {\protect\JournalTitle{Optica}} \textbf{6}, 1211--1219 (2019).

\bibitem{goy2019high}
A.~Goy, G.~Rughoobur, S.~Li, K.~Arthur, A.~I. Akinwande, and G.~Barbastathis,
  \enquote{High-resolution limited-angle phase tomography of dense layered
  objects using deep neural networks,} {\protect\JournalTitle{Proc. Natl. Acad.
  Sci.}} \textbf{116}, 19848--19856 (2019).

\bibitem{bertero1998introduction}
M.~Bertero and P.~Boccacci, \emph{Introduction to inverse problems in imaging}
  (CRC press, 1998).

\bibitem{candes2006robust}
E.~J. Cand{\`e}s, J.~Romberg, and T.~Tao, \enquote{Robust uncertainty
  principles: Exact signal reconstruction from highly incomplete frequency
  information,} {\protect\JournalTitle{IEEE Trans. Inf. Theory}} \textbf{52},
  489--509 (2006).

\bibitem{gregor2010learning}
K.~Gregor and Y.~LeCun, \enquote{Learning fast approximations of sparse
  coding,} in \emph{Proceedings of the 27th International Conference on Machine
  Learning,}  (2010), pp. 399--406.

\bibitem{barbastathis2019use}
G.~Barbastathis, A.~Ozcan, and G.~Situ, \enquote{On the use of deep learning
  for computational imaging,} {\protect\JournalTitle{Optica}} \textbf{6},
  921--943 (2019).

\bibitem{jin2017deep}
K.~H. Jin, M.~T. McCann, E.~Froustey, and M.~Unser, \enquote{Deep convolutional
  neural network for inverse problems in imaging,} {\protect\JournalTitle{IEEE
  Trans. Image Process.}} \textbf{26}, 4509--4522 (2017).

\bibitem{inv:mardani2017a}
M.~Mardani, {Enhao G}ong, J.~Y. Cheng, S.~Vasanawala, G.~Zaharchuk, M.~Alley,
  N.~Thakur, {Song Han}, W.~Daly, J.~M. Pauly, and {Lei Xing}, \enquote{Deep
  generative adversarial networks for compressed sensing automates {MRI},}
  {arXiv}:1706.00051 (2017).

\bibitem{inv:mardani2017b}
M.~Mardani, H.~Monajemi, V.~Papyan, S.~Vasanawala, D.~Donoho, and J.~Pauly,
  \enquote{Recurrent generative residual networks for proximal learning and
  automated compressive image recovery,} {arXiv}:1711.10046 (2017).

\bibitem{inv:daubechies04}
I.~Daubechies, M.~Defrise, and C.~D. Mol, \enquote{An iterative thresholding
  algorithm for linear inverse problems with a sparsity constraint,}
  {\protect\JournalTitle{Comm. Pure Appl. Math.}} \textbf{57}, 1413--1457
  (2004).

\bibitem{williams1989learning}
R.~J. Williams and D.~Zipser, \enquote{A learning algorithm for continually
  running fully recurrent neural networks,} {\protect\JournalTitle{Neural
  Comput.}} \textbf{1}, 270--280 (1989).

\bibitem{hochreiter1997long}
S.~Hochreiter and J.~Schmidhuber, \enquote{Long short-term memory,}
  {\protect\JournalTitle{Neural Comput.}} \textbf{9}, 1735--1780 (1997).

\bibitem{xingjian2015convolutional}
X.~Shi, Z.~Chen, H.~Wang, D.-Y. Yeung, W.-K. Wong, and W.-c. Woo,
  \enquote{Convolutional {LSTM} network: A machine learning approach for
  precipitation nowcasting,} in \emph{Adv. Neural Inf. Process. Syst. (NIPS),}
  (2015), pp. 802--810.

\bibitem{wang2018eidetic}
Y.~Wang, L.~Jiang, M.-H. Yang, L.-J. Li, M.~Long, and L.~Fei-Fei,
  \enquote{Eidetic {3D LSTM}: A model for video prediction and beyond,} in
  \emph{International Conference on Learning Representations (ICLR),}  (2018).

\bibitem{wang2017predrnn}
Y.~Wang, M.~Long, J.~Wang, Z.~Gao, and S.~Y. Philip, \enquote{{PredRNN}:
  Recurrent neural networks for predictive learning using spatiotemporal
  {LSTMs},} in \emph{Adv. Neural Inf. Process. Syst. (NIPS),}  (2017), pp.
  879--888.

\bibitem{wang2018predrnn++}
Y.~Wang, Z.~Gao, M.~Long, J.~Wang, and P.~S. Yu, \enquote{{PredRNN++}: Towards
  a resolution of the deep-in-time dilemma in spatiotemporal predictive
  learning,} {\protect\JournalTitle{arXiv preprint arXiv:1804.06300}}  (2018).

\bibitem{cs2018depthnet}
A.~CS~Kumar, S.~M. Bhandarkar, and M.~Prasad, \enquote{Depthnet: A recurrent
  neural network architecture for monocular depth prediction,} in \emph{Proc.
  IEEE Comput. Soc. Conf. Comput. Vis. Pattern Recognit. (CVPR) Workshops,}
  (2018), pp. 283--291.

\bibitem{wang2017shape}
W.~Wang, Q.~Huang, S.~You, C.~Yang, and U.~Neumann, \enquote{Shape inpainting
  using {3D} generative adversarial network and recurrent convolutional
  networks,} in \emph{Proceedings of the IEEE International Conference on
  Computer Vision (ICCV),}  (2017), pp. 2298--2306.

\bibitem{liu2020novel}
J.~Liu and S.~Ji, \enquote{A novel recurrent encoder-decoder structure for
  large-scale multi-view stereo reconstruction from an open aerial dataset,} in
  \emph{Proc. IEEE Comput. Soc. Conf. Comput. Vis. Pattern Recognit. (CVPR),}
  (2020), pp. 6050--6059.

\bibitem{choy20163d}
C.~B. Choy, D.~Xu, J.~Gwak, K.~Chen, and S.~Savarese, \enquote{{3D-R2N2}: A
  unified approach for single and multi-view {3D} object reconstruction,} in
  \emph{European Conference on Computer Vision (ECCV),}  (Springer, 2016), pp.
  628--644.

\bibitem{le2017multi}
T.~Le, G.~Bui, and Y.~Duan, \enquote{A multi-view recurrent neural network for
  {3D} mesh segmentation,} {\protect\JournalTitle{Comput. Graph.}} \textbf{66},
  103--112 (2017).

\bibitem{stollenga2015parallel}
M.~F. Stollenga, W.~Byeon, M.~Liwicki, and J.~Schmidhuber, \enquote{Parallel
  multi-dimensional {LSTM}, with application to fast biomedical volumetric
  image segmentation,} in \emph{Adv. Neural Inf. Process. Syst. (NIPS),}
  (2015), pp. 2998--3006.

\bibitem{hou2019multi}
Y.~Hou, J.~Kannala, and A.~Solin, \enquote{Multi-view stereo by temporal
  nonparametric fusion,} in \emph{Proceedings of the IEEE International
  Conference on Computer Vision (ICCV),}  (2019), pp. 2651--2660.

\bibitem{cho2014learning}
K.~Cho, B.~Van~Merri{\"e}nboer, C.~Gulcehre, D.~Bahdanau, F.~Bougares,
  H.~Schwenk, and Y.~Bengio, \enquote{Learning phrase representations using
  {RNN} encoder-decoder for statistical machine translation,}
  {\protect\JournalTitle{arXiv preprint arXiv:1406.1078}}  (2014).

\bibitem{goy2018low}
A.~Goy, K.~Arthur, S.~Li, and G.~Barbastathis, \enquote{Low photon count phase
  retrieval using deep learning,} {\protect\JournalTitle{Phys. Rev. Lett.}}
  \textbf{121}, 243902 (2018).

\bibitem{beck2009fast}
A.~Beck and M.~Teboulle, \enquote{Fast gradient-based algorithms for
  constrained total variation image denoising and deblurring problems,}
  {\protect\JournalTitle{IEEE Trans. Image. Process.}} \textbf{18}, 2419--2434
  (2009).

\bibitem{chambolle2004algorithm}
A.~Chambolle, \enquote{An algorithm for total variation minimization and
  applications,} {\protect\JournalTitle{J. Math. Imaging Vis.}} \textbf{20},
  89--97 (2004).

\bibitem{beck2009fista}
A.~Beck and M.~Teboulle, \enquote{A fast iterative shrinkage-thresholding
  algorithm for linear inverse problems,} {\protect\JournalTitle{SIAM J.
  Imaging Sci.}} \textbf{2}, 183--202 (2009).

\bibitem{dey2017gate}
R.~Dey and F.~M. Salemt, \enquote{Gate-variants of gated recurrent unit {(GRU)}
  neural networks,} in \emph{2017 IEEE 60th International Midwest Symposium on
  Circuits and Systems (MWSCAS),}  (IEEE, 2017), pp. 1597--1600.

\bibitem{nair2010rectified}
V.~Nair and G.~E. Hinton, \enquote{Rectified linear units improve restricted
  boltzmann machines,} in \emph{{International Conference on Machine Learning}
  (ICML),}  (2010).

\bibitem{glorot2011deep}
X.~Glorot, A.~Bordes, and Y.~Bengio, \enquote{Deep sparse rectifier neural
  networks,} in \emph{Proceedings of the fourteenth International Conference on
  Artificial Intelligence and Statistics,}  (2011), pp. 315--323.

\bibitem{sinha2017lensless}
A.~Sinha, J.~Lee, S.~Li, and G.~Barbastathis, \enquote{Lensless computational
  imaging through deep learning,} {\protect\JournalTitle{Optica}} \textbf{4},
  1117--1125 (2017).

\bibitem{gehring2016convolutional}
J.~Gehring, M.~Auli, D.~Grangier, and Y.~N. Dauphin, \enquote{A convolutional
  encoder model for neural machine translation,} {\protect\JournalTitle{arXiv
  preprint arXiv:1611.02344}}  (2016).

\bibitem{hori2017advances}
T.~Hori, S.~Watanabe, Y.~Zhang, and W.~Chan, \enquote{Advances in joint
  {CTC}-attention based end-to-end speech recognition with a deep cnn encoder
  and rnn-lm,} {\protect\JournalTitle{arXiv preprint arXiv:1706.02737}}
  (2017).

\bibitem{zhao2017learning}
R.~Zhao, R.~Yan, J.~Wang, and K.~Mao, \enquote{Learning to monitor machine
  health with convolutional bi-directional {LSTM} networks,}
  {\protect\JournalTitle{Sensors}} \textbf{17}, 273 (2017).

\bibitem{he2016deep}
K.~He, X.~Zhang, S.~Ren, and J.~Sun, \enquote{Deep residual learning for image
  recognition,} in \emph{Proc. IEEE Comput. Soc. Conf. Comput. Vis. Pattern
  Recognit. (CVPR),}  (2016), pp. 770--778.

\bibitem{ronneberger2015u}
O.~Ronneberger, P.~Fischer, and T.~Brox, \enquote{U-net: Convolutional networks
  for biomedical image segmentation,} in \emph{International Conference on
  Medical Image Computing and Computer-Assisted Intervention (MICCAI),}
  (Springer, 2015), pp. 234--241.

\bibitem{bahdanau2014neural}
D.~Bahdanau, K.~Cho, and Y.~Bengio, \enquote{Neural machine translation by
  jointly learning to align and translate,} {\protect\JournalTitle{arXiv
  preprint arXiv:1409.0473}}  (2014).

\bibitem{vaswani2017attention}
A.~Vaswani, N.~Shazeer, N.~Parmar, J.~Uszkoreit, L.~Jones, A.~N. Gomez,
  {\L}.~Kaiser, and I.~Polosukhin, \enquote{Attention is all you need,} in
  \emph{Adv. Neural Inf. Process. Syst. (NIPS),}  (2017), pp. 5998--6008.

\bibitem{li2018imaging}
S.~Li, M.~Deng, J.~Lee, A.~Sinha, and G.~Barbastathis, \enquote{Imaging through
  glass diffusers using densely connected convolutional networks,}
  {\protect\JournalTitle{Optica}} \textbf{5}, 803--813 (2018).

\bibitem{kingma2014adam}
D.~P. Kingma and J.~Ba, \enquote{Adam: A method for stochastic optimization,}
  {\protect\JournalTitle{arXiv preprint arXiv:1412.6980}}  (2014).

\bibitem{villani2003topics}
C.~Villani, \emph{Topics in optimal transportation}, 58 (American Mathematical
  Soc., 2003).

\bibitem{kolouri2017optimal}
S.~Kolouri, S.~R. Park, M.~Thorpe, D.~Slepcev, and G.~K. Rohde,
  \enquote{Optimal mass transport: Signal processing and machine-learning
  applications,} {\protect\JournalTitle{IEEE Signal Process. Mag.}}
  \textbf{34}, 43--59 (2017).

\bibitem{wang2004image}
Z.~Wang, A.~C. Bovik, H.~R. Sheikh, and E.~P. Simoncelli, \enquote{Image
  quality assessment: from error visibility to structural similarity,}
  {\protect\JournalTitle{IEEE Trans. Image Process.}} \textbf{13}, 600--612
  (2004).

\bibitem{lukovsevivcius2009reservoir}
M.~Luko{\v{s}}evi{\v{c}}ius and H.~Jaeger, \enquote{Reservoir computing
  approaches to recurrent neural network training,}
  {\protect\JournalTitle{Comput. Sci. Rev.}} \textbf{3}, 127--149 (2009).

\bibitem{lukovsevivcius2012reservoir}
M.~Luko{\v{s}}evi{\v{c}}ius, H.~Jaeger, and B.~Schrauwen, \enquote{Reservoir
  computing trends,} {\protect\JournalTitle{KI-K{\"u}nstliche Intelligenz}}
  \textbf{26}, 365--371 (2012).

\bibitem{schrauwen2007overview}
B.~Schrauwen, D.~Verstraeten, and J.~Van~Campenhout, \enquote{An overview of
  reservoir computing: theory, applications and implementations,} in
  \emph{Proceedings of the 15th European Symposium on Artificial Neural
  Networks (ESANN),}  (2007), pp. 471--482.

\end{thebibliography}
\bibliographyfullrefs{recurrent,tomography,deep_learning,algorithm,holography,inverse}

\end{document}